\documentclass[12pt,a4paper]{article}

\usepackage[utf8]{inputenc}
\usepackage[T1]{fontenc}
\usepackage{amsmath,amssymb}
\usepackage{graphicx}
\usepackage{booktabs}
\usepackage{multirow}
\usepackage{hyperref}
\usepackage[margin=2.5cm]{geometry}
\usepackage{setspace}
\usepackage[numbers,sort&compress]{natbib} 
\usepackage{xcolor}
\usepackage{caption}
\usepackage{enumitem}
\usepackage{tcolorbox}

\graphicspath{{/}{figures/}}

\title{Dose-Dependent Cardiac Complexity Changes in Children Following Prenatal Glucocorticoid Exposure: Complementary Evidence from Multiscale Entropy Analysis and ECG Foundation Models}

\author{
Nicolas B. Garnier\textsuperscript{a},
Michelle Dreiling\textsuperscript{b}
Valeska Kozik\textsuperscript{b},\\
Matthias Schwab\textsuperscript{b},
Florian Rakers\textsuperscript{b},
Martin G.\ Frasch\textsuperscript{c,d,*}
}
\date{}

\begin{document}

\maketitle

\begin{center}
\textsuperscript{a}CNRS, ENS de Lyon, LPENSL, UMR5672, 69342, Lyon cedex 07, France\\
\textsuperscript{b}Department of Neurology, Friedrich-Schiller-University of Jena, Jena, Germany\\
\textsuperscript{c}Institute on Human Development and Disability, University of Washington, Seattle, WA, USA\\
\textsuperscript{d}Health Stream Analytics, LLC\\[1em]
\textsuperscript{*}Corresponding author: Martin G.\ Frasch, Institute on Human Development and Disability, University of Washington, Seattle, WA 98195, USA.\\
E-mail: \href{mailto:mfrasch@uw.edu}{mfrasch@uw.edu}
\end{center}

\vspace{1em}

\begin{abstract}
\noindent\textbf{Background}
Prenatal glucocorticoid exposure alters cardiac development, but whether persistent cardiac effects in childhood follow a dose-response relationship remains unknown. We recently showed that ECG foundation models detect robust cardiac differences between steroid-exposed and control children, while traditional heart rate variability metrics lose significance after covariate adjustment. Here, we investigate the dose-response dimension using complementary analytical approaches.

\noindent\textbf{Methods}
We studied 49 children (ages 8--15) whose mothers received betamethasone during pregnancy for multiple sclerosis: 12 low-dose ({$<$}5\,g cumulative), 13 high-dose ({$\geq$}5\,g), and 24 controls. Five-minute ECG recordings during the Trier Social Stress Test yielded 251 observations. We computed 12 multiscale complexity features and tested 11 ECG foundation model (FM) dimensions using linear mixed models, Kruskal--Wallis tests with Dunn's post-hoc comparisons, Spearman correlations, and Jonckheere--Terpstra trend tests.

\noindent\textbf{Findings}
The binary exposed-versus-controls comparison showed no significant complexity effects ($p>0.39$). However, dose-based analysis revealed that high-dose children exhibited significantly faster entropy rate ($h$) decay rates than low-dose children ($p=0.031$); neither sample entropy nor approximate entropy decay rates reached significance ($p=0.18$ and $p=0.12$, respectively). Effects localized to the mental arithmetic stress segment (Kruskal--Wallis $p=0.005$; Dunn's $p=0.004$). A cross-condition robustness analysis confirmed that $h$ decay rate is invariant to input signal choice and normalization ($r>0.98$), while sample and approximate entropy are not. In contrast, the 11 FM dimensions showed weak dose-response evidence: only 1 of 22 covariate-adjusted contrasts survived FDR correction, with paradoxically stronger low-dose effects.

\noindent\textbf{Interpretation}
The entropy rate decay rate---uniquely robust across input conditions---reveals a dose-dependent effect on cardiac autonomic dynamics under cognitive stress, while FM dimensions detect a dose-independent morphological ``exposure fingerprint.'' These exploratory findings suggest a two-component model of prenatal glucocorticoid cardiac programming ---~morphological (dose-independent) and dynamical (dose-dependent)~--- providing more complete characterization than either approach alone. Given the small sample size, these results should be considered hypothesis-generating and require replication in larger cohorts.
\end{abstract}

\vspace{1em}

\begin{tcolorbox}[colback=gray!5, colframe=gray!50, title={\textbf{Research in Context}}]

\textbf{Evidence before this study}\\
We searched PubMed and Google Scholar for studies published up to January 2026 using the terms ``prenatal glucocorticoid,'' ``cardiac programming,'' ``ECG,'' ``heart rate variability,'' ``multiscale entropy,'' and ``dose-response.'' Animal studies demonstrate that antenatal corticosteroids alter cardiac structure, autonomic receptor expression, and blood pressure regulation. Human studies report associations between prenatal glucocorticoid exposure and cardiovascular changes in offspring, but results are inconsistent and no study has examined dose-response relationships for cardiac complexity or ECG morphology features. Our companion preprint showed that ECG foundation models detect persistent cardiac differences between steroid-exposed and control children, but treated the exposed group as monolithic despite a 12-fold range in cumulative betamethasone dose.

\textbf{Added value of this study}\\
This study is the first to examine dose-response relationships for prenatal glucocorticoid cardiac effects using both multiscale entropy analysis and ECG foundation models. We demonstrate that the entropy rate ($h$) decay rate---quantifying how rapidly cardiac dynamical complexity diminishes beyond its peak timescale---shows a clear dose-dependent pattern localized to the peak cognitive stress segment. Crucially, a cross-condition robustness analysis confirms this metric is invariant to input signal choice (HR vs.\ RRI) and normalization ($r>0.98$), while sample and approximate entropy decay rates are not. Foundation model dimensions capture a dose-independent morphological signal. This dissociation reveals a two-component model of cardiac programming that was invisible when treating exposed subjects as a single group.

\textbf{Implications of all the available evidence}\\
These findings establish that prenatal glucocorticoid cardiac effects have both morphological and dynamical components with distinct dose-response profiles. Clinical monitoring of exposed children may benefit from combining ECG morphology analysis (to detect any exposure) with complexity assessment under stress (to estimate dose-dependent autonomic impact). The results also have methodological implications: studies treating glucocorticoid-exposed groups as monolithic risk false-negative conclusions by masking real dose-dependent effects. Replication in larger cohorts is needed before clinical translation.
\end{tcolorbox}

\vspace{1em}

\newpage

\section{Introduction}

Prenatal glucocorticoid administration is among the most common pharmacological interventions in obstetrics, used to accelerate fetal lung maturation in threatened preterm birth and as treatment for maternal autoimmune conditions including multiple sclerosis \cite{roberts2017antenatal,alwan2013reproductive}. While short-term benefits are well established, growing evidence suggests that prenatal glucocorticoid exposure may program lasting cardiovascular alterations \cite{barker2007origins,gluckman2008effect}. Animal studies demonstrate that antenatal corticosteroids alter cardiac structure, cardiomyocyte calcium handling, autonomic receptor expression, and blood pressure regulation \cite{rogzielinska2013glucocorticoid,segar1997effect,derks1997comparative}. Human studies report associations with elevated blood pressure, altered heart rate, and modified stress reactivity in exposed offspring, though results are inconsistent \cite{kelly2012antenatal,doyle2000antenatal,alexander2012impact}.

We recently reported that deep learning-based ECG foundation models detect persistent cardiac differences between steroid-exposed and control children at age 8, while traditional heart rate variability (HRV) metrics lose significance after covariate adjustment \cite{frasch2026longterm}. Specifically, 11 of 512 foundation model embedding dimensions showed significant group differences (FDR-corrected $p<0.05$, $|$Cohen's $d| = 0.79$--$1.72$), likely capturing subtle ECG waveform morphology features invisible to traditional HRV analysis \cite{frasch2026longterm}.

A critical unaddressed question is whether these cardiac effects follow a dose-response relationship. Cumulative betamethasone doses in our cohort ranged from 1.25\,g to 15.0\,g---a 12-fold range that could mask dose-dependent effects when treated as a monolithic exposure. Dose-response evidence is essential for mechanistic understanding and clinical risk stratification.

Multiscale entropy analysis offers a complementary lens for cardiac assessment. Unlike traditional HRV metrics, multiscale complexity measures quantify information content and predictability across temporal scales \cite{costa2002multiscale,richman2000physiological}, capturing fractal-like regulatory structure that time-domain and frequency-domain metrics miss \cite{goldberger2000physiobank,peng1995quantification}.

Here, we investigate dose-response using two complementary frameworks: (1) 12 multiscale complexity features derived from entropy rate, approximate entropy, and sample entropy, and (2) the 11 ECG foundation model dimensions identified previously \cite{frasch2026longterm}. By comparing how these feature types respond to steroid dose, we aim to elucidate the nature of cardiac programming and assess complementary biomarker potential.

\section{Methods}

\subsection{Study population}

The cohort has been described previously \cite{frasch2026longterm}. Forty-nine children (ages 8--15 years) were recruited from Jena University Hospital and Ruhr University Bochum / St.\ Josef Hospital between October 2020 and August 2023. Children were born to mothers who received methylprednisolone during pregnancy for multiple sclerosis (exposed, $n=25$) or were unexposed controls ($n=24$). The study was approved by ethics committees at both sites (UKJ: 2020-1668-3-BO; RUB: 21-7192 BR) and registered at ClinicalTrials.gov (NCT04832269). Informed consent was obtained from all participants and their legal guardians.

Exposed subjects were categorized by cumulative methylprednisolone exposure: low-dose ({$<$}5\,g, $n=12$; range 1.25--3.0\,g) and high-dose ({$\geq$}5\,g, $n=13$; range 5.0--15.0\,g). The 5\,g threshold was chosen to approximate the median cumulative dose and to produce approximately balanced dose groups. To complement this categorical approach, we also tested continuous dose-response using Spearman correlations with cumulative dose (Sum\_Kort) among exposed subjects. Groups were balanced for sex ($\chi^2$ $p=0.54$) and nearly balanced for gestational age (controls: $38.9 \pm 1.4$ weeks; exposed: $38.1 \pm 1.8$ weeks; $p=0.09$) \cite{frasch2026longterm}.

\subsection{ECG recording protocol}

Holter ECG recordings were obtained during the Trier Social Stress Test (TSST) \cite{kirschbaum1993trier}, segmented into: baseline, preparation, speech anticipation, mental arithmetic, early recovery, late recovery, and extended recovery. Each segment comprised approximately 5 minutes of continuous single-channel ECG at 250\,Hz. The 49 subjects contributed 251 total observations (5--7 segments per subject).

\subsection{Multiscale complexity features}

Twelve (3$\times$4) complexity features were computed from RR interval (RRI) time series normalized by their standard deviation, derived from each 5-minute ECG segment as follows. RR intervals were extracted using ensemble R-peak detection with signal quality index (SQI) filtering, as described previously \cite{frasch2026longterm}, and validated for physiological range (330--1500\,ms, corresponding to heart rates of 40--180\,bpm). Normalization by standard deviation ensures that entropy measures reflect temporal structure (complexity) rather than amplitude variability, and renders the input complementary to the foundation model features which operate on raw ECG morphology. A cross-condition robustness analysis (Section~\ref{sec:robustness}) confirmed that the primary finding---dose-dependent $h$ decay rate---is invariant to input signal choice (heart rate vs.\ RRI) and normalization.

\begin{figure}[htbp]
\centering
\includegraphics[width=0.75\textwidth]{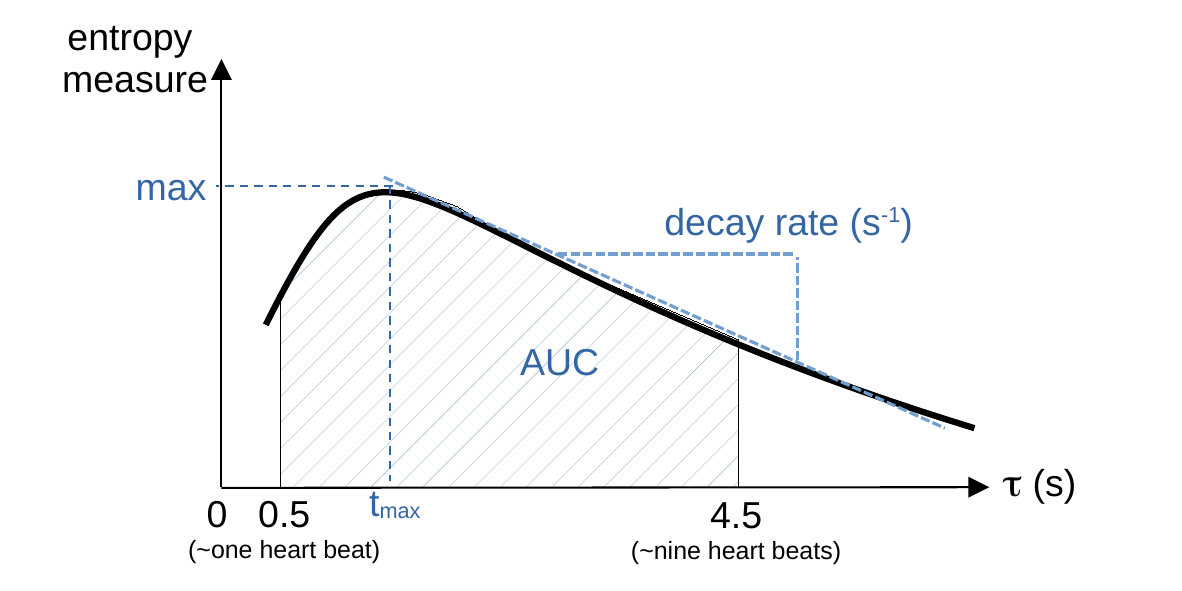}
\caption{Schematics of the four summary statistics used on each complexity measure. "max" is the value of maximum over all time-scales and "$t_{\rm max}$" its corresponding time-scale. "AUC" is the area under the curve over the interval [0,5 - 4.5] seconds and "decay rate" is the slope of the linear decrease of the complexity measure above $t_{\rm max}$.}
\label{fig:schematics_definitions}
\end{figure}

Three multiscale entropy measures were computed: entropy rate ($h$)~\cite{Granero:2017}, approximate entropy (ApEn) \cite{pincus1991approximate}, and sample entropy (SampEn) \cite{lake2002sample}. Entropy rate quantifies the rate of information production; ApEn and SampEn quantify regularity and predictability. For ApEn and SampEn, the embedding dimension was $m=2$ and the tolerance was $r = 0.2 \times \text{SD}(x)$, where SD$(x)$ is the standard deviation of the original RRI series; $r$ was held fixed across scales to ensure commensurability \cite{richman2000physiological}. SampEn excludes self-matches, providing a less biased estimate than ApEn for finite time series \cite{lake2002sample,Granero:2017}.

Each measure was computed across temporal scales from 0.5 to 7\,seconds using a stride-based embedding procedure with time-step d$\tau = 0.04$\,s, yielding scale-dependent entropy curves. Four summary statistics were extracted per measure (Figure~\ref{fig:schematics_definitions}): area under the curve (AUC; average complexity across scales), maximum value (max; peak complexity), timescale at maximum ($t_{\rm max}$ in seconds), and post-peak decay rate (dec\_rate, s$^{-1}$; quantifying how rapidly complexity diminishes beyond its peak timescale). Analyses used smoothed data (d$\tau$=0.04\,s with local averaging over 5 consecutive timescale values) as the primary dataset, with unsmoothed data (d$\tau$=0.04\,s without averaging) as a robustness check. Results were consistent across both preprocessing conditions; where both are reported, the difference serves as an empirical error estimate.

\subsection{ECG foundation model features}

Foundation model features were extracted using an Advanced Foundation CNN generating 512-dimensional embeddings from raw ECG segments \cite{frasch2026longterm}. We focused on the 11 dimensions surviving covariate-adjusted multiple testing in the binary comparison (FDR $p<0.05$, $|d|>0.8$): dimensions 255, 367, 52, 318, 220, 172, 437, 493, 51, 402, and 353.

\subsection{Statistical analysis}

For each feature, linear mixed models (LMMs) were fitted with group (binary or three-level dose), sex, gestational age, and age as fixed effects and a random intercept per subject \cite{seabold2010statsmodels}. Kruskal--Wallis tests compared dose groups per segment, with Dunn's post-hoc pairwise comparisons (Bonferroni-corrected) \cite{dunn1964multiple}. Jonckheere--Terpstra tests assessed monotonic dose-response trends \cite{jonckheere1954distribution}. Spearman correlations tested continuous dose-response among exposed subjects ($n=25$). Benjamini--Hochberg FDR correction was applied within each analysis. Binary and three-class classification used logistic regression, random forest, SVM, and XGBoost with stratified group 5-fold cross-validation.

\section{Results}

\subsection{Binary exposure comparison reveals no complexity effects}

LMMs with binary exposure showed no significant effect on any complexity feature (all $p>0.39$; Table~\ref{tab:binary}), with negligible effect sizes. This null finding was consistent across preprocessing conditions.

\begin{table}[htbp]
\centering
\caption{Linear mixed models: binary exposure effect on complexity features (selected). Smoothed data shown as primary; unsmoothed values as robustness check. Input: RRI normalized by standard deviation.}
\label{tab:binary}
\begin{tabular}{lcccc}
\toprule
Feature & Coef (smoothed) & $p$ & Coef (unsmoothed) & $p$ \\
\midrule
h\_AUC & $-0.013$ & 0.740 & $-0.013$ & 0.740 \\
h\_dec\_rate & $-0.006$ & 0.389 & $-0.006$ & 0.387 \\
AE\_dec\_rate & $-0.003$ & 0.702 & $-0.003$ & 0.721 \\
SE\_AUC & $-0.021$ & 0.509 & $-0.021$ & 0.509 \\
SE\_dec\_rate & $-0.002$ & 0.825 & $-0.003$ & 0.719 \\
\bottomrule
\end{tabular}
\vspace{0.3em}

{\footnotesize All 12 features shown in Supplementary Table~S1. No feature reached $p<0.39$.}
\end{table}

\subsection{Dose-response analysis unmasks complexity effects}

The three-level dose grouping revealed effects entirely masked in the binary comparison. High-dose subjects showed significantly faster entropy rate ($h$) decay rates than low-dose subjects (Table~\ref{tab:dose}): coef=$+0.018$, $p=0.031$ (smoothed); $p=0.026$ (unsmoothed). Neither SampEn decay (coef=$+0.015$, $p=0.18$) nor ApEn decay (coef=$+0.015$, $p=0.12$) reached significance. Control-versus-low-dose contrasts were also significant for $h$ decay ($p=0.049$, smoothed; $p=0.045$, unsmoothed).

\begin{table}[htbp]
\centering
\caption{Linear mixed models: dose-based effects on complexity decay rates. Input: RRI normalized by SD. Smoothed data shown as primary; unsmoothed as robustness check.}
\label{tab:dose}
\begin{tabular}{llcccc}
\toprule
Feature & Contrast & Coef (smth) & $p$ (smth) & Coef (unsmth) & $p$ (unsmth) \\
\midrule
h\_dec\_rate & High vs.\ Low & $+0.018$ & \textbf{0.031} & $+0.019$ & \textbf{0.026} \\
SE\_dec\_rate & High vs.\ Low & $+0.015$ & 0.177 & $+0.016$ & 0.133 \\
AE\_dec\_rate & High vs.\ Low & $+0.015$ & 0.124 & $+0.015$ & 0.121 \\
h\_dec\_rate & Ctrl vs.\ Low & $+0.015$ & \textbf{0.049} & $+0.016$ & \textbf{0.045} \\
SE\_dec\_rate & Ctrl vs.\ Low & $+0.009$ & 0.343 & $+0.011$ & 0.250 \\
AE\_dec\_rate & Ctrl vs.\ Low & $+0.010$ & 0.233 & $+0.010$ & 0.239 \\
\bottomrule
\end{tabular}
\end{table}

\subsection{Dose effects localize to the mental arithmetic stress segment}

Per-segment Kruskal--Wallis tests localized the dose effect to the mental arithmetic segment (Table~\ref{tab:m3m4_kw}). Only h\_dec\_rate showed a significant three-group difference ($H=10.69$, $p=0.005$, $\eta^2=0.20$, FDR $q=0.033$); neither SE\_dec\_rate ($H=4.49$, $p=0.11$) nor any other complexity feature reached significance. Dunn's post-hoc comparisons confirmed the low-dose versus high-dose contrast drove the effect ($p_\text{Bonf}=0.004$), with a marginal low-dose versus control difference ($p_\text{Bonf}=0.075$). No other segment showed FDR-significant effects. Unsmoothed data yielded concordant results ($H=10.35$, $p=0.006$, FDR $q=0.040$; Dunn's low-vs-high $p_\text{Bonf}=0.005$).

\begin{table}[htbp]
\centering
\caption{Kruskal--Wallis results and Dunn's post-hoc: mental arithmetic segment. Input: RRI normalized by SD.}
\label{tab:m3m4_kw}
\begin{tabular}{lcccc}
\toprule
Feature & $H$ & $p$ & $\eta^2$ & Dunn's Low vs.\ High ($p_\text{Bonf}$) \\
\midrule
h\_dec\_rate & 10.69 & \textbf{0.005} & 0.20 & \textbf{0.004} \\
SE\_dec\_rate & 4.49 & 0.106 & 0.06 & --- \\
AE\_dec\_rate & --- & --- & --- & --- \\
\bottomrule
\end{tabular}
\end{table}

\subsection{Robustness analysis: $h$ decay rate is uniquely invariant}
\label{sec:robustness}

To assess the dependence of our findings on input signal choice, we repeated the entire complexity analysis under four conditions: heart rate (HR) or RR intervals (RRI), each raw or normalized by standard deviation. Table~\ref{tab:robustness_corr} shows cross-condition Pearson correlations ($N=251$ segments) for the three decay rate metrics.

The $h$ decay rate showed near-perfect agreement across all conditions ($r=0.983$--$0.998$), confirming that this metric is virtually invariant to signal type and normalization. In contrast, SampEn decay rate showed highly variable correlations ($r=0.31$--$0.95$), and ApEn decay rate was essentially uncorrelated between raw and normalized variants ($r=-0.06$ to $0.98$). This invariance is expected theoretically: $h$ is computed via $k$-nearest-neighbor mutual information estimation, which depends on rank-order distances and is therefore invariant under monotonic transformations---precisely the relationship between HR and RRI ($\text{HR}=60/\text{RRI}$) and between raw and normalized signals (division by a constant).

Critically, the dose-response significance was also robust: Kruskal--Wallis tests on the mental arithmetic segment showed h\_dec\_rate reaching $p<0.01$ in all four conditions (Table~\ref{tab:robustness_kw_main}), while SE\_dec\_rate was significant only in the HR-raw condition ($p=0.002$) and lost significance in all other conditions ($p>0.10$). Low-dose versus high-dose Mann--Whitney comparisons confirmed consistent large effects for h\_dec\_rate (Cohen's $d=1.53$--$1.66$, $p<0.005$) across all four conditions.

\begin{table}[htbp]
\centering
\caption{Cross-condition Pearson correlations for decay rates ($N=251$ segments). Only $h$ decay rate shows near-perfect agreement ($r>0.98$) across all input conditions.}
\label{tab:robustness_corr}
\small
\begin{tabular}{l cc cc cc}
\toprule
& \multicolumn{2}{c}{$h$ decay rate} & \multicolumn{2}{c}{SampEn decay rate} & \multicolumn{2}{c}{ApEn decay rate} \\
\cmidrule(lr){2-3} \cmidrule(lr){4-5} \cmidrule(lr){6-7}
Pair & $r$ & & $r$ & & $r$ & \\
\midrule
HR\textsubscript{raw} vs HR\textsubscript{norm}   & \textbf{0.997} & & 0.713 & & 0.091 \\
HR\textsubscript{raw} vs RRI\textsubscript{raw}   & \textbf{0.983} & & 0.502 & & 0.099 \\
HR\textsubscript{raw} vs RRI\textsubscript{norm}  & \textbf{0.985} & & 0.698 & & 0.089 \\
HR\textsubscript{norm} vs RRI\textsubscript{raw}  & \textbf{0.985} & & 0.316 & & $-0.045$ \\
HR\textsubscript{norm} vs RRI\textsubscript{norm} & \textbf{0.986} & & 0.953 & & 0.980 \\
RRI\textsubscript{raw} vs RRI\textsubscript{norm} & \textbf{0.998} & & 0.310 & & $-0.060$ \\
\midrule
Range & \textbf{0.983--0.998} & & 0.310--0.953 & & $-0.060$--0.980 \\
\bottomrule
\end{tabular}
\end{table}

\begin{table}[htbp]
\centering
\caption{Dose-response Kruskal--Wallis tests ($H$, $p$) on the mental arithmetic segment across four input conditions. Only $h$ decay rate is significant in all conditions.}
\label{tab:robustness_kw_main}
\small
\begin{tabular}{l cc cc cc}
\toprule
& \multicolumn{2}{c}{$h$ decay rate} & \multicolumn{2}{c}{SampEn decay rate} & \multicolumn{2}{c}{ApEn decay rate} \\
\cmidrule(lr){2-3} \cmidrule(lr){4-5} \cmidrule(lr){6-7}
Condition & $H$ & $p$ & $H$ & $p$ & $H$ & $p$ \\
\midrule
HR\textsubscript{raw}  & 10.16 & \textbf{0.006} & 12.49 & \textbf{0.002} & 1.90 & 0.387 \\
HR\textsubscript{norm} & 10.59 & \textbf{0.005} & 4.09  & 0.129            & 4.41 & 0.110 \\
RRI\textsubscript{raw} & 10.18 & \textbf{0.006} & 4.51  & 0.105            & 0.37 & 0.832 \\
RRI\textsubscript{norm}& 10.69 & \textbf{0.005} & 4.49  & 0.106            & 4.51 & 0.105 \\
\bottomrule
\end{tabular}
\end{table}

\subsection{Foundation model dimensions show weak dose-response}

In contrast, the 11 FM dimensions showed limited dose-response evidence. Kruskal--Wallis tests (88 tests: 8 segments $\times$ 11 dimensions) yielded no FDR-significant results. Dunn's post-hoc comparisons revealed five Bonferroni-significant contrasts, all involving control-versus-dose-group differences rather than dose-gradient differences (Table~\ref{tab:fm_dunns}).

\begin{table}[htbp]
\centering
\caption{Significant Dunn's post-hoc contrasts for FM dimensions.}
\label{tab:fm_dunns}
\begin{tabular}{llccc}
\toprule
Segment & FM Dimension & Contrast & $z$ & $p_\text{Bonf}$ \\
\midrule
Late Recovery & FM\_Dim367 & Ctrl vs.\ Low & 3.11 & \textbf{0.006} \\
Speech Anticipation & FM\_Dim255 & Ctrl vs.\ High & 2.62 & \textbf{0.027} \\
Late Recovery & FM\_Dim353 & Ctrl vs.\ Low & 2.46 & \textbf{0.041} \\
Speech Anticipation & FM\_Dim318 & Ctrl vs.\ High & 2.48 & \textbf{0.039} \\
Late Recovery & FM\_Dim255 & Ctrl vs.\ High & 2.43 & \textbf{0.045} \\
\bottomrule
\end{tabular}
\end{table}

Covariate-adjusted LMMs yielded one FDR-significant result: FM\_Dim367 low-dose versus controls ($d=-1.07$, FDR $q=0.036$). Notably, low-dose effects were consistently stronger than high-dose effects (8 of 11 dimensions), contrary to a dose-response gradient (Table~\ref{tab:fm_lmm}). Spearman correlations with continuous dose showed no significant associations (all FDR $q>0.85$). Jonckheere--Terpstra tests found no monotonic trends (all FDR $q>0.58$).

\begin{table}[htbp]
\centering
\caption{Covariate-adjusted LMM: FM dimensions by dose group.}
\label{tab:fm_lmm}
\begin{tabular}{llccc}
\toprule
FM Dimension & Contrast & Cohen's $d$ & $p$ & FDR $q$ \\
\midrule
FM\_Dim367 & Low vs.\ Ctrl & $-1.07$ & 0.003 & \textbf{0.036} \\
FM\_Dim51 & Low vs.\ Ctrl & $-0.93$ & 0.014 & 0.078 \\
FM\_Dim255 & Low vs.\ Ctrl & $-0.83$ & 0.032 & 0.118 \\
FM\_Dim367 & High vs.\ Ctrl & $-0.57$ & 0.110 & 0.575 \\
FM\_Dim51 & High vs.\ Ctrl & $-0.65$ & 0.083 & 0.575 \\
FM\_Dim255 & High vs.\ Ctrl & $-0.53$ & 0.162 & 0.575 \\
\bottomrule
\end{tabular}
\end{table}

\subsection{Head-to-head comparison confirms complementarity}

\begin{figure}[htbp]
\centering
\includegraphics[width=\textwidth]{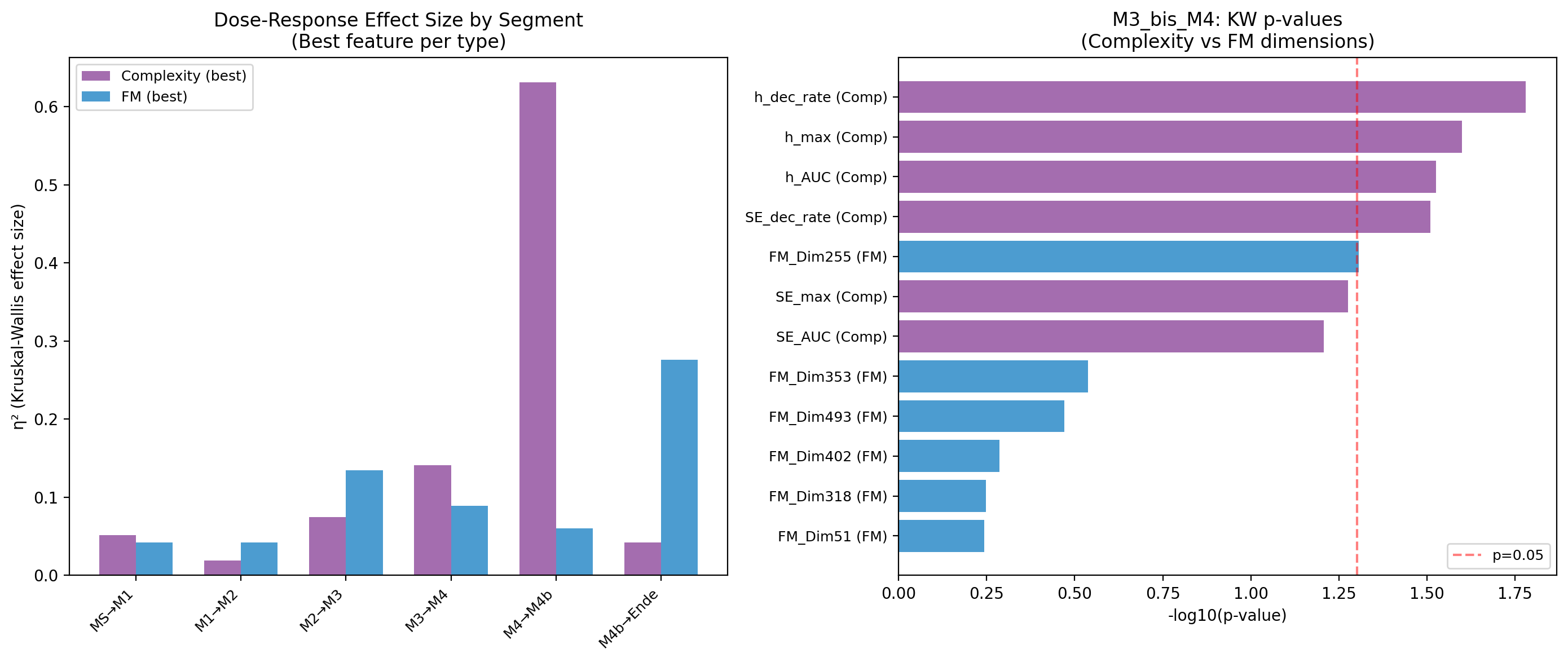}
\caption{Comparative Kruskal--Wallis effect sizes ($\eta^2$) for FM dimensions versus complexity features across TSST recording segments. Complexity features show stronger dose-response in the mental arithmetic segment, while FM dimensions dominate in recovery segments with exposure-based (not dose-gradient) effects.}
\label{fig:comparison}
\end{figure}

Direct comparison of Kruskal--Wallis effect sizes revealed a striking dissociation (Table~\ref{tab:comparison}; Fig.~\ref{fig:comparison}). In the mental arithmetic segment, complexity outperformed FM (h\_dec\_rate $\eta^2=0.20$ vs.\ FM\_Dim255 $\eta^2=0.09$), with only complexity showing significant dose-gradient contrasts (Dunn's $p=0.004$). In late recovery, FM substantially outperformed complexity (FM\_Dim367 $\eta^2=0.28$ vs.\ h\_max $\eta^2=0.04$), but FM contrasts were exposure-based, not dose-gradient.

\begin{table}[htbp]
\centering
\caption{Kruskal--Wallis effect sizes: FM vs.\ complexity by segment.}
\label{tab:comparison}
\begin{tabular}{llclcl}
\toprule
Segment & Best FM & FM $\eta^2$ & Best Complexity & Compl.\ $\eta^2$ & Dose-response? \\
\midrule
Mental Arith. & FM\_Dim255 & 0.09 & h\_dec\_rate & \textbf{0.20} & Complexity only \\
Late Recovery & FM\_Dim367 & \textbf{0.28} & h\_max & 0.04 & Neither (FM: exp.) \\
Speech Antic. & FM\_Dim255 & 0.13 & h\_AUC & 0.08 & Neither \\
Early Recovery & FM\_Dim353 & 0.06 & h\_dec\_rate & \textbf{0.58} & Complexity trend \\
Baseline & FM\_Dim367 & 0.04 & h\_dec\_rate & 0.04 & Neither \\
Ext.\ Recovery & FM\_Dim318 & \textbf{0.30} & SE\_max & 0.08 & Neither \\
\bottomrule
\end{tabular}
\end{table}

\subsection{Machine learning classification at chance level}

All classifiers performed at or below chance for both binary (chance=0.50) and three-class (chance=0.33) classification using complexity features (Supplementary Table~S5). The best binary performance was XGBoost (accuracy=0.45, AUC=0.44). This poor classification performance likely reflects the fundamental nature of the complexity signal: dose effects are concentrated in a single segment (mental arithmetic) and in only two of twelve features (decay rates), providing insufficient discriminative information when aggregated across all segments and features. Furthermore, the small sample size ($n=49$) relative to the number of features, combined with the high inter-individual variability inherent in cardiac complexity measures, limits the ability of supervised classifiers to learn generalizable decision boundaries. We note that the companion study achieved similarly modest classification performance using FM dimensions alone \cite{frasch2026longterm}, suggesting that the challenge is inherent to the sample size rather than the feature type. Future work could explore targeted feature selection focused on the mental arithmetic segment---where dose effects concentrate---combined with dimensionality reduction techniques (e.g., PCA or autoencoders) to extract maximally discriminative low-dimensional representations before classification.

\section{Discussion}

This study reveals a fundamental dissociation in how two complementary ECG analysis approaches capture prenatal glucocorticoid cardiac effects. Complexity decay rates show a clear dose-dependent pattern concentrated in the peak cognitive stress phase, while FM dimensions detect a dose-independent signal with paradoxically stronger low-dose effects.

\subsection{The binary comparison masks real dose-dependent effects}

The null binary comparison (all $p>0.38$), despite a real dose-dependent signal, has a straightforward explanation: pooling low-dose and high-dose subjects inflates within-group variance. The dose groups show opposing patterns relative to controls, creating a bimodal distribution that attenuates mean differences. This has important methodological implications---studies treating glucocorticoid-exposed groups as monolithic may systematically miss dose-dependent effects, and dose stratification should be standard practice when cumulative exposure data are available.

\subsection{Entropy rate decay rate as a uniquely robust dose-dependent biomarker}

The entropy rate ($h$) decay rate emerged as the sole robust complexity biomarker: it was the only decay rate metric to (a) show significant dose-response in the LMM ($p=0.031$), (b) reach significance in the per-segment Kruskal--Wallis test ($p=0.005$, FDR $q=0.033$), and (c) maintain significance across all four input conditions in the robustness analysis ($p<0.006$ in every condition; Table~\ref{tab:robustness_kw_main}).

The robustness of $h$ decay rate has a principled mathematical explanation: it is computed via $k$-nearest-neighbor mutual information estimation, where distances are rank-based and therefore invariant under monotonic transformations. Since HR $=60/$RRI is monotonic, and normalization by standard deviation is likewise monotonic, $h$ decay rate is theoretically expected to be invariant to these choices---as confirmed empirically ($r>0.98$ across all condition pairs; Table~\ref{tab:robustness_corr}). In contrast, SampEn and ApEn use template matching with an absolute tolerance threshold ($r \times \text{SD}$), making them sensitive to the distributional geometry of the input signal.

SampEn decay rate, which appeared significant in the previous HR-raw analysis ($p=0.002$), lost significance when computed from RRI-normalized input ($p=0.11$). This input-condition dependence disqualifies it as a reliable biomarker and underscores the importance of robustness verification---a finding with implications for the broader multiscale entropy literature.

Higher $h$ decay rates in the high-dose group indicate a narrower bandwidth of complex cardiac dynamics, consistent with dose-dependent perturbation of multiscale autonomic architecture. This may reflect reduced autonomic reserve limiting complex dynamics under cognitive stress \cite{thayer2009claude}, simplified regulatory hierarchy preserving peak complexity but reducing temporal breadth \cite{ivanov1999multifractality}, or altered sympathovagal coupling at longer timescales \cite{segar1997effect,derks1997comparative}.

The localization to the mental arithmetic segment is consistent with the stress-reactivity hypothesis \cite{entringer2015prenatal}: under maximal cognitive demand, dose-dependent differences in autonomic capacity become apparent. This converges with our prior finding that baseline ECG was insufficient for group discrimination \cite{frasch2026longterm}.

\subsection{A two-component model of cardiac programming}

Together, our findings suggest a two-component model:
\begin{enumerate}[nosep]
    \item \textbf{Morphological component} (FM): Structural and electrical myocardial changes triggered during a critical developmental window, producing a dose-independent ``exposure fingerprint'' detectable in ECG waveform morphology. The paradoxical pattern of stronger low-dose effects supports a threshold model where timing matters more than total dose \cite{sloboda2005synthetic,matthews2000antenatal}.
    \item \textbf{Dynamical component} (entropy rate decay): Functional autonomic changes scaling with cumulative dose, manifest as altered complexity dynamics under stress. Specifically, the entropy rate decay rate---the uniquely robust complexity biomarker identified through cross-condition analysis---captures how rapidly cardiac dynamical complexity diminishes beyond its peak timescale. This is biologically plausible: while morphological changes may be triggered at any exposure level, autonomic functional alteration may scale with total dose through effects on nerve fiber density, receptor expression, and neural-cardiac coupling \cite{rogzielinska2013glucocorticoid,segar1997effect}.
\end{enumerate}

This model explains the central puzzle of our prior work---why FM dimensions detect steroid exposure while HRV metrics do not \cite{frasch2026longterm}. Traditional HRV captures autonomic modulation but uses metrics sensitive to demographic confounders and insufficiently sensitive to dose-dependent complexity changes.

\subsection{Interpreting foundation model dimensions}

Although the 11 discriminative FM dimensions were identified through a data-driven approach without explicit feature engineering, their dose-independent behavior and large effect sizes ($|d|=0.79$--$1.72$) constrain their likely physiological substrate. The foundation model operates on raw ECG waveforms, suggesting sensitivity to morphological features such as QRS complex shape, ST-segment geometry, T-wave amplitude and duration, and inter-wave timing intervals. Prenatal glucocorticoid exposure is known to alter cardiomyocyte structure, gap junction distribution, and ion channel expression \cite{rogzielinska2013glucocorticoid}, all of which would manifest as subtle but consistent ECG waveform changes. The dose-independent nature of these effects is consistent with a ``critical window'' programming model: glucocorticoid-induced structural changes may be triggered during a sensitive developmental period regardless of total dose, producing a persistent morphological fingerprint \cite{sloboda2005synthetic,matthews2000antenatal}. Future work should systematically apply gradient-based attribution methods---including integrated gradients, saliency maps, and layer-wise relevance propagation---to identify which ECG waveform segments drive each discriminative dimension, potentially linking specific FM dimensions to QRS morphology, repolarization patterns, or conduction velocity changes. Preliminary investigations suggest that such methods are feasible for the foundation model architecture used here; systematic application across all 11 dimensions and recording segments would be essential for clinical translation and mechanistic understanding.

\subsection{Clinical implications and limitations}

The complementary biomarker framework suggests that clinical assessment could combine FM-based screening (binary exposure detection) with complexity-based dose estimation under standardized stress.

However, several important limitations must be acknowledged. First, the small sample size ($n=49$ total; $n=12$--$13$ per dose group) substantially limits statistical power and generalizability. Post-hoc power calculations indicate that with $n=12$--$13$ per dose group and $\alpha=0.05$ (two-tailed), we had approximately 80\% power to detect large effects ($d \geq 1.2$) but only ${\sim}40$\% power for medium effects ($d=0.7$). The significant finding for h\_dec\_rate ($p=0.031$) should be considered hypothesis-generating until replicated in larger cohorts; a well-powered replication targeting $d=0.8$ would require approximately 25--30 subjects per dose group.

Second, the dose categorization at 5\,g cumulative betamethasone, while approximately median-splitting the exposed group and producing balanced subgroups, remains pragmatic. The concordance between categorical (Kruskal--Wallis) and continuous (Spearman) analyses partially addresses this concern---if the threshold were grossly misplaced, we would expect continuous dose measures to outperform categorical ones, which was not observed. Nevertheless, future studies with larger samples could explore data-driven threshold identification or spline-based dose-response modeling.

Third, the observational design and the specific clinical context of maternal multiple sclerosis (MS) treatment introduce potential confounders. Maternal MS itself may affect offspring cardiac development through shared genetic susceptibility, altered intrauterine inflammatory environment, or co-administered medications. We controlled for sex, gestational age, and age at examination in the LMM analyses, but unmeasured confounders---including maternal disease severity, concurrent medications, timing of steroid exposure during gestation, socioeconomic factors, and childhood physical activity levels---could influence the results. The inclusion of a control group from the same clinical centers partially mitigates selection bias, but residual confounding cannot be excluded in this observational design.

Fourth, while the robustness analysis demonstrates that h\_dec\_rate is invariant to input signal choice and normalization with current parameters ($m=2$, $k=5$), sensitivity to embedding dimension and $k$-NN parameter values was not tested; future work should explore whether the dose-response signal persists across parameter ranges. Fifth, complexity features depend on R-peak detection quality, and while our ensemble detection approach with SQI filtering provides robustness \cite{frasch2026longterm}, segments with lower signal quality could introduce noise that differentially affects dose groups. Sixth, the biological interpretation of FM dimensions remains indirect; characterization of the specific ECG morphological features captured by these dimensions through attribution analyses is an important next step.

Despite these limitations, and while findings from this exploratory study warrant cautious interpretation, the convergent evidence from multiple complementary statistical approaches (LMM, Kruskal--Wallis, Dunn's post-hoc) and the specificity of effects to a physiologically plausible segment (peak cognitive stress) provide confidence that the dose-response signal is genuine. These hypothesis-generating results require replication in independent cohorts with larger sample sizes, diverse clinical indications for prenatal glucocorticoid use, and prospective longitudinal designs.


\section*{Contributors}

NGa and MGF conceived the analytical approach and designed the study. MGF performed the computational analyses, including foundation model feature extraction, statistical modelling, and machine learning classification. NGa performed multiscale entropy computation, developed and applied the multiscale entropy methodology and approved the statistical interpretation. MS and FR designed and conducted the original clinical study, recruited participants, and collected ECG data. MGF wrote the first draft of the manuscript. All authors reviewed, edited, and approved the final manuscript. MGF and NGa directly accessed and verified the underlying data.

\section*{Data sharing statement}

Analysis code is available at \url{https://github.com/martinfrasch/Florian_Nicolas_ER}. De-identified data are available upon reasonable request with IRB approval. The companion preprint is available at \url{https://doi.org/10.64898/2026.02.02.26345391}.

\section*{Declaration of interests}

M. Schwab received funding for travel or speaker honoraria and has served on advisory boards for Janssen, Almirall, Bayer Healthcare, Biogen, BMS, Sanofi-Genzyme, Merck Healthcare, Novartis, Roche, HEXAL AG, and TEVA; and received research support from Novartis and Bayer Healthcare. F. Rakers has received research grants from Merck KGaA and Biogen GmbH; speaker honoraria from Merck KGaA, Biogen GmbH, BMS GmbH, Roche GmbH, and Novartis GmbH; and travel funding from Merck KGaA and Biogen GmbH. M. Frasch holds patents on maternal and fetal monitoring and equity in pregnancy health start-ups. All other authors declare no competing interests.

\section*{Acknowledgements and funding}

The original clinical study was supported by the Grant for Multiple Sclerosis Innovation 2020 (Merck Healthcare KGaA). The computational analysis in the present study received no specific funding. The authors thank the families who participated in this study.

\section*{Ethics statement}

This study was approved by the institutional ethics committees at Jena University Hospital (UKJ Reference: 2020-1668-3-BO) and Ruhr University Bochum (RUB Reference: 21-7192 BR) and registered at ClinicalTrials.gov (NCT04832269). Written informed consent was obtained from all participants and their legal guardians.

\bibliographystyle{unsrtnat}
\bibliography{references}

\begin{thebibliography}{27}
\providecommand{\natexlab}[1]{#1}
\providecommand{\url}[1]{\texttt{#1}}
\expandafter\ifx\csname urlstyle\endcsname\relax
  \providecommand{\doi}[1]{doi: #1}\else
  \providecommand{\doi}{doi: \begingroup \urlstyle{rm}\Url}\fi

\bibitem[Roberts et~al.(2017)Roberts, Brown, Medley, and
  Dalziel]{roberts2017antenatal}
Devender Roberts, Julie Brown, Nancy Medley, and Stuart~R Dalziel.
\newblock Antenatal corticosteroids for accelerating fetal lung maturation for
  women at risk of preterm birth.
\newblock \emph{Cochrane Database of Systematic Reviews}, 3:\penalty0 CD004454,
  2017.

\bibitem[Alwan et~al.(2013)Alwan, Yee, Dybalski, et~al.]{alwan2013reproductive}
Sadika Alwan, Irene~M Yee, Marisa Dybalski, et~al.
\newblock Reproductive decision making after the diagnosis of multiple
  sclerosis ({MS}).
\newblock \emph{Multiple Sclerosis Journal}, 19\penalty0 (3):\penalty0
  351--358, 2013.

\bibitem[Barker(2007)]{barker2007origins}
David~JP Barker.
\newblock The origins of the developmental origins theory.
\newblock \emph{Journal of Internal Medicine}, 261\penalty0 (5):\penalty0
  412--417, 2007.

\bibitem[Gluckman et~al.(2008)Gluckman, Hanson, Cooper, and
  Thornburg]{gluckman2008effect}
Peter~D Gluckman, Mark~A Hanson, Cyrus Cooper, and Kent~L Thornburg.
\newblock Effect of in utero and early-life conditions on adult health and
  disease.
\newblock \emph{New England Journal of Medicine}, 359\penalty0 (1):\penalty0
  61--73, 2008.

\bibitem[Rog-Zielinska et~al.(2013)Rog-Zielinska, Thomson, Sheridan,
  et~al.]{rogzielinska2013glucocorticoid}
Eva~A Rog-Zielinska, Amber Thomson, Hayley Sheridan, et~al.
\newblock Glucocorticoid receptor is required for foetal heart maturation.
\newblock \emph{Human Molecular Genetics}, 22\penalty0 (16):\penalty0
  3269--3282, 2013.

\bibitem[Segar et~al.(1997)Segar, Scholz, Bedell, Smith, Huss, and
  Guillery]{segar1997effect}
Jeffrey~L Segar, Timothy~D Scholz, Karen~A Bedell, Ola~M Smith, Daniel~J Huss,
  and Edward~N Guillery.
\newblock Effect of cortisol on gene expression of the renin-angiotensin system
  in fetal sheep.
\newblock \emph{Pediatric Research}, 41\penalty0 (2):\penalty0 251--256, 1997.

\bibitem[Derks et~al.(1997)Derks, Giussani, Jenkins,
  et~al.]{derks1997comparative}
Jan~B Derks, Dino~A Giussani, Sandra~L Jenkins, et~al.
\newblock A comparative study of cardiovascular, endocrine and behavioural
  effects of betamethasone and dexamethasone administration to fetal sheep.
\newblock \emph{The Journal of Physiology}, 499\penalty0 (Pt 1):\penalty0
  217--226, 1997.

\bibitem[Kelly et~al.(2012)Kelly, Lewandowski, Worton,
  et~al.]{kelly2012antenatal}
Brenda~A Kelly, Adam~J Lewandowski, Sarah~A Worton, et~al.
\newblock Antenatal glucocorticoid exposure and long-term alterations in aortic
  function and glucose metabolism.
\newblock \emph{Pediatrics}, 129\penalty0 (5):\penalty0 e1282--e1290, 2012.

\bibitem[Doyle et~al.(2000)Doyle, Ford, Davis, and
  Callanan]{doyle2000antenatal}
Lex~W Doyle, Gary~W Ford, Noni~M Davis, and Catherine Callanan.
\newblock Antenatal corticosteroid therapy and blood pressure at 14 years of
  age in preterm children.
\newblock \emph{Clinical Science}, 98\penalty0 (2):\penalty0 137--142, 2000.

\bibitem[Alexander et~al.(2012)Alexander, Rosenl{\"o}cher, Stalder,
  et~al.]{alexander2012impact}
Nina Alexander, Franziska Rosenl{\"o}cher, Tobias Stalder, et~al.
\newblock Impact of antenatal synthetic glucocorticoid exposure on endocrine
  stress reactivity in term-born children.
\newblock \emph{The Journal of Clinical Endocrinology \& Metabolism},
  97\penalty0 (10):\penalty0 3538--3544, 2012.

\bibitem[Frasch et~al.(2026)Frasch, Schwab, and Rakers]{frasch2026longterm}
Martin~G Frasch, Matthias Schwab, and Florian Rakers.
\newblock Long-term cardiac autonomic effects of prenatal steroid exposure: a
  machine learning approach integrating heart rate variability and {ECG}
  foundation models.
\newblock \emph{medRxiv}, 2026.
\newblock \doi{10.64898/2026.02.02.26345391}.
\newblock Preprint.

\bibitem[Costa et~al.(2002)Costa, Goldberger, and Peng]{costa2002multiscale}
Madalena Costa, Ary~L Goldberger, and C-K Peng.
\newblock Multiscale entropy analysis of complex physiologic time series.
\newblock \emph{Physical Review Letters}, 89\penalty0 (6):\penalty0 068102,
  2002.

\bibitem[Richman and Moorman(2000)]{richman2000physiological}
Joshua~S Richman and J~Randall Moorman.
\newblock Physiological time-series analysis using approximate entropy and
  sample entropy.
\newblock \emph{American Journal of Physiology--Heart and Circulatory
  Physiology}, 278\penalty0 (6):\penalty0 H2039--H2049, 2000.

\bibitem[Goldberger et~al.(2000)Goldberger, Amaral, Glass,
  et~al.]{goldberger2000physiobank}
Ary~L Goldberger, Luis A~N Amaral, Leon Glass, et~al.
\newblock {PhysioBank}, {PhysioToolkit}, and {PhysioNet}: components of a new
  research resource for complex physiologic signals.
\newblock \emph{Circulation}, 101\penalty0 (23):\penalty0 e215--e220, 2000.

\bibitem[Peng et~al.(1995)Peng, Havlin, Stanley, and
  Goldberger]{peng1995quantification}
C-K Peng, Shlomo Havlin, H~Eugene Stanley, and Ary~L Goldberger.
\newblock Quantification of scaling exponents and crossover phenomena in
  nonstationary heartbeat time series.
\newblock \emph{Chaos}, 5\penalty0 (1):\penalty0 82--87, 1995.

\bibitem[Kirschbaum et~al.(1993)Kirschbaum, Pirke, and
  Hellhammer]{kirschbaum1993trier}
Clemens Kirschbaum, Karl-Martin Pirke, and Dirk~H Hellhammer.
\newblock The `{T}rier {S}ocial {S}tress {T}est' -- a tool for investigating
  psychobiological stress responses in a laboratory setting.
\newblock \emph{Neuropsychobiology}, 28\penalty0 (1--2):\penalty0 76--81, 1993.

\bibitem[Granero-Belinchon et~al.(2017)Granero-Belinchon, Roux, Abry, Doret,
  and Garnier]{Granero:2017}
C.~Granero-Belinchon, S.~G. Roux, P.~Abry, M.~Doret, and N.~B. Garnier.
\newblock Information theory to probe intrapartum fetal heart rate dynamics.
\newblock \emph{Entropy}, 19:\penalty0 640, 2017.
\newblock \doi{10.3390/e19120640}.

\bibitem[Pincus(1991)]{pincus1991approximate}
Steven~M Pincus.
\newblock Approximate entropy as a measure of system complexity.
\newblock \emph{Proceedings of the National Academy of Sciences}, 88\penalty0
  (6):\penalty0 2297--2301, 1991.

\bibitem[Lake et~al.(2002)Lake, Richman, Griffin, and Moorman]{lake2002sample}
Douglas~E Lake, Joshua~S Richman, M~Pamela Griffin, and J~Randall Moorman.
\newblock Sample entropy analysis of neonatal heart rate variability.
\newblock \emph{American Journal of Physiology--Regulatory, Integrative and
  Comparative Physiology}, 283\penalty0 (3):\penalty0 R789--R797, 2002.

\bibitem[Seabold and Perktold(2010)]{seabold2010statsmodels}
Skipper Seabold and Josef Perktold.
\newblock Statsmodels: econometric and statistical modeling with {P}ython.
\newblock In \emph{Proceedings of the 9th Python in Science Conference}, 2010.

\bibitem[Dunn(1964)]{dunn1964multiple}
Olive~Jean Dunn.
\newblock Multiple comparisons using rank sums.
\newblock \emph{Technometrics}, 6\penalty0 (3):\penalty0 241--252, 1964.

\bibitem[Jonckheere(1954)]{jonckheere1954distribution}
A~R Jonckheere.
\newblock A distribution-free $k$-sample test against ordered alternatives.
\newblock \emph{Biometrika}, 41\penalty0 (1/2):\penalty0 133--145, 1954.

\bibitem[Thayer and Lane(2009)]{thayer2009claude}
Julian~F Thayer and Richard~D Lane.
\newblock Claude {B}ernard and the heart--brain connection: further elaboration
  of a model of neurovisceral integration.
\newblock \emph{Neuroscience \& Biobehavioral Reviews}, 33\penalty0
  (2):\penalty0 81--88, 2009.

\bibitem[Ivanov et~al.(1999)Ivanov, Amaral, Goldberger,
  et~al.]{ivanov1999multifractality}
Plamen~Ch Ivanov, Luis A~Nunes Amaral, Ary~L Goldberger, et~al.
\newblock Multifractality in human heartbeat dynamics.
\newblock \emph{Nature}, 399\penalty0 (6735):\penalty0 461--465, 1999.

\bibitem[Entringer et~al.(2015)Entringer, Buss, and
  Wadhwa]{entringer2015prenatal}
Sonja Entringer, Claudia Buss, and Pathik~D Wadhwa.
\newblock Prenatal stress, development, health and disease risk: a
  psychobiological perspective -- 2015 {C}urt {R}ichter {A}ward paper.
\newblock \emph{Psychoneuroendocrinology}, 62:\penalty0 366--375, 2015.

\bibitem[Sloboda et~al.(2005)Sloboda, Challis, Moss, and
  Newnham]{sloboda2005synthetic}
Deborah~M Sloboda, John R~G Challis, Timothy J~M Moss, and John~P Newnham.
\newblock Synthetic glucocorticoids: antenatal administration and long-term
  implications.
\newblock \emph{Current Pharmaceutical Design}, 11\penalty0 (11):\penalty0
  1459--1472, 2005.

\bibitem[Matthews(2000)]{matthews2000antenatal}
Stephen~G Matthews.
\newblock Antenatal glucocorticoids and programming of the developing {CNS}.
\newblock \emph{Pediatric Research}, 47\penalty0 (3):\penalty0 291--300, 2000.

\end{thebibliography}

\newpage
\appendix
\section*{Supplementary Material}

\begin{table}[htbp]
\centering
\caption{Complete linear mixed model results for all 12 complexity features. Model: feature $\sim$ group + Sex + gestational\_age + age + (1$|$subject). Input: RRI normalized by SD. Smoothed (S) data shown as primary; unsmoothed (U) as robustness check.}
\label{tab:s1}
{\footnotesize
\textbf{(A) Binary exposure (Group\_exp: exposed vs.\ control)}\\[0.3em]
\begin{tabular}{l cc cc}
\toprule
Feature & Coef (S) & $p$ (S) & Coef (U) & $p$ (U) \\
\midrule
h\_AUC & $-0.013$ & 0.74 & $-0.013$ & 0.74 \\
h\_max & $-0.012$ & 0.80 & $-0.009$ & 0.85 \\
h\_$t_{\rm max}$ & $-0.055$ & 0.56 & $-0.057$ & 0.56 \\
h\_dec\_rate & $-0.006$ & 0.39 & $-0.006$ & 0.39 \\
AE\_AUC & $-0.010$ & 0.70 & $-0.010$ & 0.70 \\
AE\_max & $-0.010$ & 0.75 & $-0.010$ & 0.77 \\
AE\_$t_{\rm max}$ & $+0.119$ & 0.45 & $+0.156$ & 0.30 \\
AE\_dec\_rate & $-0.003$ & 0.70 & $-0.003$ & 0.72 \\
SE\_AUC & $-0.021$ & 0.51 & $-0.021$ & 0.51 \\
SE\_max & $-0.020$ & 0.58 & $-0.019$ & 0.61 \\
SE\_$t_{\rm max}$ & $+0.034$ & 0.87 & $+0.055$ & 0.76 \\
SE\_dec\_rate & $-0.002$ & 0.83 & $-0.003$ & 0.72 \\
\bottomrule
\end{tabular}
\vspace{1em}

\textbf{(B) Three-level dose (Group\_k: high vs.\ low [T.2] and control vs.\ low [T.3])}\\[0.3em]
\begin{tabular}{ll cc cc}
\toprule
Feature & Contrast & Coef (S) & $p$ (S) & Coef (U) & $p$ (U) \\
\midrule
h\_AUC & High vs.\ Low & $+0.014$ & 0.79 & $+0.014$ & 0.79 \\
h\_AUC & Ctrl vs.\ Low & $+0.021$ & 0.67 & $+0.021$ & 0.67 \\
\addlinespace[2pt]
h\_max & High vs.\ Low & $+0.023$ & 0.71 & $+0.021$ & 0.73 \\
h\_max & Ctrl vs.\ Low & $+0.023$ & 0.67 & $+0.019$ & 0.73 \\
\addlinespace[2pt]
h\_$t_{\rm max}$ & High vs.\ Low & $+0.093$ & 0.46 & $+0.121$ & 0.35 \\
h\_$t_{\rm max}$ & Ctrl vs.\ Low & $+0.102$ & 0.37 & $+0.119$ & 0.31 \\
\addlinespace[2pt]
h\_dec\_rate & High vs.\ Low & $+0.018$ & \textbf{0.031} & $+0.019$ & \textbf{0.026} \\
h\_dec\_rate & Ctrl vs.\ Low & $+0.015$ & \textbf{0.049} & $+0.016$ & \textbf{0.045} \\
\addlinespace[2pt]
AE\_AUC & High vs.\ Low & $+0.021$ & 0.55 & $+0.021$ & 0.55 \\
AE\_AUC & Ctrl vs.\ Low & $+0.021$ & 0.52 & $+0.021$ & 0.52 \\
\addlinespace[2pt]
AE\_max & High vs.\ Low & $+0.037$ & 0.39 & $+0.037$ & 0.38 \\
AE\_max & Ctrl vs.\ Low & $+0.029$ & 0.45 & $+0.029$ & 0.46 \\
\addlinespace[2pt]
AE\_$t_{\rm max}$ & High vs.\ Low & $-0.125$ & 0.56 & $-0.115$ & 0.58 \\
AE\_$t_{\rm max}$ & Ctrl vs.\ Low & $-0.183$ & 0.34 & $-0.215$ & 0.25 \\
\addlinespace[2pt]
AE\_dec\_rate & High vs.\ Low & $+0.015$ & 0.12 & $+0.015$ & 0.12 \\
AE\_dec\_rate & Ctrl vs.\ Low & $+0.010$ & 0.23 & $+0.010$ & 0.24 \\
\addlinespace[2pt]
SE\_AUC & High vs.\ Low & $+0.004$ & 0.93 & $+0.004$ & 0.93 \\
SE\_AUC & Ctrl vs.\ Low & $+0.023$ & 0.56 & $+0.023$ & 0.56 \\
\addlinespace[2pt]
SE\_max & High vs.\ Low & $+0.023$ & 0.63 & $+0.024$ & 0.62 \\
SE\_max & Ctrl vs.\ Low & $+0.032$ & 0.47 & $+0.031$ & 0.48 \\
\addlinespace[2pt]
SE\_$t_{\rm max}$ & High vs.\ Low & $-0.258$ & 0.38 & $-0.128$ & 0.60 \\
SE\_$t_{\rm max}$ & Ctrl vs.\ Low & $-0.166$ & 0.53 & $-0.121$ & 0.58 \\
\addlinespace[2pt]
SE\_dec\_rate & High vs.\ Low & $+0.015$ & 0.18 & $+0.016$ & 0.13 \\
SE\_dec\_rate & Ctrl vs.\ Low & $+0.009$ & 0.34 & $+0.011$ & 0.25 \\
\bottomrule
\end{tabular}
}
\end{table}

\clearpage

\begin{table}[htbp]
\centering
\caption{Per-segment Kruskal--Wallis tests for complexity features by three-level dose group (smoothed data). FDR correction applied within each segment. Unsmoothed results are concordant and available in the data repository.}
\label{tab:s2}
{\scriptsize
\begin{tabular}{ll rr rr l}
\toprule
Feature & Segment & $n$ & $H$ & $p$ & $\eta^2$ & FDR $q$ \\
\midrule
h\_AUC & Baseline & 49 & 0.54 & 0.76 & 0.000 & 0.97 \\
h\_AUC & Preparation & 48 & 0.06 & 0.97 & 0.000 & 0.97 \\
h\_AUC & Speech Antic. & 47 & 1.16 & 0.56 & 0.000 & 0.97 \\
h\_AUC & Mental Arith. & 47 & 0.84 & 0.66 & 0.000 & 0.97 \\
h\_AUC & Early Recovery & 11 & 0.21 & 0.90 & 0.000 & 0.97 \\
h\_AUC & Late Recovery & 32 & 0.11 & 0.94 & 0.000 & 0.97 \\
h\_AUC & Ext.\ Recovery & 15 & 3.45 & 0.18 & 0.121 & 0.97 \\
\addlinespace[3pt]
h\_max & Baseline & 49 & 0.25 & 0.88 & 0.000 & 0.91 \\
h\_max & Preparation & 48 & 0.19 & 0.91 & 0.000 & 0.91 \\
h\_max & Speech Antic. & 47 & 1.00 & 0.61 & 0.000 & 0.91 \\
h\_max & Mental Arith. & 47 & 0.93 & 0.63 & 0.000 & 0.91 \\
h\_max & Early Recovery & 11 & 1.30 & 0.52 & 0.000 & 0.91 \\
h\_max & Late Recovery & 32 & 0.27 & 0.87 & 0.000 & 0.91 \\
h\_max & Ext.\ Recovery & 15 & 3.17 & 0.20 & 0.097 & 0.91 \\
\addlinespace[3pt]
h\_$t_{\rm max}$ & Baseline & 49 & 0.36 & 0.84 & 0.000 & 0.87 \\
h\_$t_{\rm max}$ & Preparation & 48 & 0.81 & 0.67 & 0.000 & 0.87 \\
h\_$t_{\rm max}$ & Speech Antic. & 47 & 0.40 & 0.82 & 0.000 & 0.87 \\
h\_$t_{\rm max}$ & Mental Arith. & 47 & 1.82 & 0.40 & 0.000 & 0.87 \\
h\_$t_{\rm max}$ & Early Recovery & 11 & 3.05 & 0.22 & 0.131 & 0.87 \\
h\_$t_{\rm max}$ & Late Recovery & 32 & 1.13 & 0.57 & 0.000 & 0.87 \\
h\_$t_{\rm max}$ & Ext.\ Recovery & 15 & 0.29 & 0.87 & 0.000 & 0.87 \\
\addlinespace[3pt]
h\_dec\_rate & Baseline & 49 & 3.18 & 0.20 & 0.026 & 0.48 \\
h\_dec\_rate & Preparation & 48 & 0.75 & 0.69 & 0.000 & 0.86 \\
h\_dec\_rate & Speech Antic. & 47 & 0.33 & 0.85 & 0.000 & 0.86 \\
h\_dec\_rate & Mental Arith. & 47 & 10.69 & \textbf{0.005} & 0.197 & \textbf{0.033} \\
h\_dec\_rate & Early Recovery & 11 & 6.66 & \textbf{0.036} & 0.582 & 0.13 \\
h\_dec\_rate & Late Recovery & 32 & 0.31 & 0.86 & 0.000 & 0.86 \\
h\_dec\_rate & Ext.\ Recovery & 15 & 1.07 & 0.59 & 0.000 & 0.86 \\
\addlinespace[3pt]
SE\_AUC & Baseline & 49 & 0.13 & 0.94 & 0.000 & 1.00 \\
SE\_AUC & Preparation & 48 & 0.01 & 1.00 & 0.000 & 1.00 \\
SE\_AUC & Speech Antic. & 47 & 0.78 & 0.68 & 0.000 & 1.00 \\
SE\_AUC & Mental Arith. & 47 & 1.18 & 0.55 & 0.000 & 1.00 \\
SE\_AUC & Early Recovery & 11 & 1.50 & 0.47 & 0.000 & 1.00 \\
SE\_AUC & Late Recovery & 32 & 0.61 & 0.74 & 0.000 & 1.00 \\
SE\_AUC & Ext.\ Recovery & 15 & 5.46 & 0.065 & 0.288 & 0.46 \\
\addlinespace[3pt]
SE\_max & Baseline & 49 & 0.43 & 0.81 & 0.000 & 0.81 \\
SE\_max & Preparation & 48 & 0.42 & 0.81 & 0.000 & 0.81 \\
SE\_max & Speech Antic. & 47 & 1.88 & 0.39 & 0.000 & 0.81 \\
SE\_max & Mental Arith. & 47 & 1.98 & 0.37 & 0.000 & 0.81 \\
SE\_max & Early Recovery & 11 & 1.33 & 0.52 & 0.000 & 0.81 \\
SE\_max & Late Recovery & 32 & 0.69 & 0.71 & 0.000 & 0.81 \\
SE\_max & Ext.\ Recovery & 15 & 2.32 & 0.31 & 0.027 & 0.81 \\
\addlinespace[3pt]
SE\_$t_{\rm max}$ & Baseline & 49 & 0.12 & 0.94 & 0.000 & 0.96 \\
SE\_$t_{\rm max}$ & Preparation & 48 & 0.61 & 0.74 & 0.000 & 0.96 \\
SE\_$t_{\rm max}$ & Speech Antic. & 47 & 0.08 & 0.96 & 0.000 & 0.96 \\
SE\_$t_{\rm max}$ & Mental Arith. & 47 & 0.83 & 0.66 & 0.000 & 0.96 \\
SE\_$t_{\rm max}$ & Early Recovery & 11 & 5.33 & 0.070 & 0.416 & 0.49 \\
SE\_$t_{\rm max}$ & Late Recovery & 32 & 0.23 & 0.89 & 0.000 & 0.96 \\
SE\_$t_{\rm max}$ & Ext.\ Recovery & 15 & 2.04 & 0.36 & 0.003 & 0.96 \\
\addlinespace[3pt]
SE\_dec\_rate & Baseline & 49 & 0.82 & 0.66 & 0.000 & 0.73 \\
SE\_dec\_rate & Preparation & 48 & 0.81 & 0.67 & 0.000 & 0.73 \\
SE\_dec\_rate & Speech Antic. & 47 & 1.29 & 0.52 & 0.000 & 0.73 \\
SE\_dec\_rate & Mental Arith. & 47 & 4.49 & 0.11 & 0.057 & 0.73 \\
SE\_dec\_rate & Early Recovery & 11 & 1.33 & 0.52 & 0.000 & 0.73 \\
SE\_dec\_rate & Late Recovery & 32 & 1.34 & 0.51 & 0.000 & 0.73 \\
SE\_dec\_rate & Ext.\ Recovery & 15 & 0.62 & 0.73 & 0.000 & 0.73 \\
\bottomrule
\end{tabular}
}
\end{table}

\clearpage

\begin{table}[htbp]
\centering
\caption{Per-segment Mann--Whitney $U$ tests for complexity features by binary exposure (smoothed data). FDR correction applied within each feature across segments. Unsmoothed results are concordant and available in the data repository.}
\label{tab:s3}
{\scriptsize
\begin{tabular}{ll rr rr r}
\toprule
Feature & Segment & $n_1$/$n_2$ & $U$ & $p$ & $r$ & FDR $q$ \\
\midrule
h\_AUC & Baseline & 24/25 & 264 & 0.48 & +0.120 & 0.84 \\
h\_AUC & Preparation & 24/24 & 293 & 0.93 & $-$0.017 & 0.93 \\
h\_AUC & Speech Antic. & 23/24 & 229 & 0.32 & +0.170 & 0.84 \\
h\_AUC & Mental Arith. & 22/25 & 317 & 0.38 & $-$0.153 & 0.84 \\
h\_AUC & Early Recovery & 4/7 & 16 & 0.79 & $-$0.143 & 0.92 \\
h\_AUC & Late Recovery & 15/17 & 135 & 0.79 & $-$0.059 & 0.92 \\
h\_AUC & Ext.\ Recovery & 7/8 & 44 & 0.072 & $-$0.571 & 0.50 \\
\addlinespace[3pt]
h\_max & Baseline & 24/25 & 275 & 0.62 & +0.083 & 0.93 \\
h\_max & Preparation & 24/24 & 293 & 0.93 & $-$0.017 & 0.93 \\
h\_max & Speech Antic. & 23/24 & 232 & 0.35 & +0.159 & 0.93 \\
h\_max & Mental Arith. & 22/25 & 285 & 0.84 & $-$0.036 & 0.93 \\
h\_max & Early Recovery & 4/7 & 17 & 0.65 & $-$0.214 & 0.93 \\
h\_max & Late Recovery & 15/17 & 131 & 0.91 & $-$0.027 & 0.93 \\
h\_max & Ext.\ Recovery & 7/8 & 43 & 0.094 & $-$0.536 & 0.66 \\
\addlinespace[3pt]
h\_$t_{\rm max}$ & Baseline & 24/25 & 270 & 0.56 & +0.098 & 0.75 \\
h\_$t_{\rm max}$ & Preparation & 24/24 & 331 & 0.38 & $-$0.149 & 0.75 \\
h\_$t_{\rm max}$ & Speech Antic. & 23/24 & 306 & 0.54 & $-$0.107 & 0.75 \\
h\_$t_{\rm max}$ & Mental Arith. & 22/25 & 324 & 0.30 & $-$0.178 & 0.75 \\
h\_$t_{\rm max}$ & Early Recovery & 4/7 & 18 & 0.57 & $-$0.250 & 0.75 \\
h\_$t_{\rm max}$ & Late Recovery & 15/17 & 132 & 0.86 & $-$0.039 & 0.86 \\
h\_$t_{\rm max}$ & Ext.\ Recovery & 7/8 & 32 & 0.64 & $-$0.161 & 0.75 \\
\addlinespace[3pt]
h\_dec\_rate & Baseline & 24/25 & 268 & 0.53 & +0.107 & 0.91 \\
h\_dec\_rate & Preparation & 24/24 & 302 & 0.78 & $-$0.049 & 0.91 \\
h\_dec\_rate & Speech Antic. & 23/24 & 300 & 0.62 & $-$0.087 & 0.91 \\
h\_dec\_rate & Mental Arith. & 22/25 & 296 & 0.66 & $-$0.076 & 0.91 \\
h\_dec\_rate & Early Recovery & 4/7 & 18 & 0.53 & $-$0.286 & 0.91 \\
h\_dec\_rate & Late Recovery & 15/17 & 139 & 0.68 & $-$0.090 & 0.91 \\
h\_dec\_rate & Ext.\ Recovery & 7/8 & 29 & 0.96 & $-$0.036 & 0.96 \\
\addlinespace[3pt]
SE\_AUC & Baseline & 24/25 & 285 & 0.77 & +0.050 & 0.90 \\
SE\_AUC & Preparation & 24/24 & 285 & 0.96 & +0.010 & 0.96 \\
SE\_AUC & Speech Antic. & 23/24 & 249 & 0.57 & +0.098 & 0.90 \\
SE\_AUC & Mental Arith. & 22/25 & 321 & 0.33 & $-$0.167 & 0.90 \\
SE\_AUC & Early Recovery & 4/7 & 17 & 0.65 & $-$0.214 & 0.90 \\
SE\_AUC & Late Recovery & 15/17 & 141 & 0.62 & $-$0.106 & 0.90 \\
SE\_AUC & Ext.\ Recovery & 7/8 & 48 & \textbf{0.021} & $-$0.714 & 0.14 \\
\addlinespace[3pt]
SE\_max & Baseline & 24/25 & 288 & 0.82 & +0.040 & 0.94 \\
SE\_max & Preparation & 24/24 & 292 & 0.94 & $-$0.014 & 0.94 \\
SE\_max & Speech Antic. & 23/24 & 234 & 0.38 & +0.152 & 0.70 \\
SE\_max & Mental Arith. & 22/25 & 319 & 0.35 & $-$0.160 & 0.70 \\
SE\_max & Early Recovery & 4/7 & 19 & 0.41 & $-$0.357 & 0.70 \\
SE\_max & Late Recovery & 15/17 & 146 & 0.50 & $-$0.145 & 0.70 \\
SE\_max & Ext.\ Recovery & 7/8 & 41 & 0.15 & $-$0.464 & 0.70 \\
\addlinespace[3pt]
SE\_$t_{\rm max}$ & Baseline & 24/25 & 314 & 0.79 & $-$0.045 & 0.92 \\
SE\_$t_{\rm max}$ & Preparation & 24/24 & 263 & 0.61 & +0.087 & 0.92 \\
SE\_$t_{\rm max}$ & Speech Antic. & 23/24 & 271 & 0.92 & +0.018 & 0.92 \\
SE\_$t_{\rm max}$ & Mental Arith. & 22/25 & 316 & 0.39 & $-$0.147 & 0.92 \\
SE\_$t_{\rm max}$ & Early Recovery & 4/7 & 22 & 0.16 & $-$0.571 & 0.71 \\
SE\_$t_{\rm max}$ & Late Recovery & 15/17 & 136 & 0.78 & $-$0.063 & 0.92 \\
SE\_$t_{\rm max}$ & Ext.\ Recovery & 7/8 & 16 & 0.20 & +0.411 & 0.71 \\
\addlinespace[3pt]
SE\_dec\_rate & Baseline & 24/25 & 312 & 0.82 & $-$0.040 & 1.00 \\
SE\_dec\_rate & Preparation & 24/24 & 275 & 0.80 & +0.045 & 1.00 \\
SE\_dec\_rate & Speech Antic. & 23/24 & 235 & 0.39 & +0.149 & 1.00 \\
SE\_dec\_rate & Mental Arith. & 22/25 & 296 & 0.66 & $-$0.076 & 1.00 \\
SE\_dec\_rate & Early Recovery & 4/7 & 14 & 1.00 & +0.000 & 1.00 \\
SE\_dec\_rate & Late Recovery & 15/17 & 157 & 0.27 & $-$0.231 & 1.00 \\
SE\_dec\_rate & Ext.\ Recovery & 7/8 & 28 & 1.00 & +0.000 & 1.00 \\
\bottomrule
\end{tabular}
}
\end{table}

\clearpage

\begin{table}[htbp]
\centering
\caption{Complete FM dimension dose-response results. (A) Covariate-adjusted LMM: FM\_dim $\sim$ Gk\_low + Gk\_high + Sex + gest\_alter + (1$|$subject). (B) Spearman correlations with cumulative dose (Sum\_Kort) among exposed subjects ($n=25$). (C) Jonckheere--Terpstra trend tests (control $<$ low $<$ high).}
\label{tab:s4}
{\footnotesize
\textbf{(A) Covariate-adjusted linear mixed models}\\[0.3em]
\begin{tabular}{ll rrr r}
\toprule
Dimension & Contrast & Cohen's $d$ & $p$ & FDR $q$ \\
\midrule
FM\_Dim367 & Low vs.\ Ctrl & -1.07 & \textbf{0.003} & \textbf{0.036} \\
FM\_Dim51 & Low vs.\ Ctrl & -0.93 & \textbf{0.014} & 0.078 \\
FM\_Dim255 & Low vs.\ Ctrl & -0.83 & \textbf{0.032} & 0.12 \\
FM\_Dim402 & Low vs.\ Ctrl & -0.65 & 0.069 & 0.17 \\
FM\_Dim353 & Low vs.\ Ctrl & -0.68 & 0.076 & 0.17 \\
FM\_Dim51 & High vs.\ Ctrl & -0.65 & 0.083 & 0.58 \\
FM\_Dim367 & High vs.\ Ctrl & -0.57 & 0.11 & 0.58 \\
FM\_Dim255 & High vs.\ Ctrl & -0.53 & 0.16 & 0.58 \\
FM\_Dim437 & Low vs.\ Ctrl & -0.43 & 0.22 & 0.40 \\
FM\_Dim402 & High vs.\ Ctrl & -0.41 & 0.24 & 0.58 \\
FM\_Dim437 & High vs.\ Ctrl & -0.38 & 0.26 & 0.58 \\
FM\_Dim353 & High vs.\ Ctrl & -0.37 & 0.33 & 0.58 \\
FM\_Dim493 & High vs.\ Ctrl & 0.22 & 0.43 & 0.58 \\
FM\_Dim52 & High vs.\ Ctrl & 0.27 & 0.47 & 0.58 \\
FM\_Dim318 & High vs.\ Ctrl & -0.28 & 0.49 & 0.58 \\
FM\_Dim318 & Low vs.\ Ctrl & -0.28 & 0.50 & 0.74 \\
FM\_Dim220 & High vs.\ Ctrl & 0.23 & 0.53 & 0.58 \\
FM\_Dim220 & Low vs.\ Ctrl & 0.21 & 0.58 & 0.74 \\
FM\_Dim493 & Low vs.\ Ctrl & -0.14 & 0.63 & 0.74 \\
FM\_Dim52 & Low vs.\ Ctrl & 0.16 & 0.67 & 0.74 \\
FM\_Dim172 & High vs.\ Ctrl & 0.13 & 0.72 & 0.72 \\
FM\_Dim172 & Low vs.\ Ctrl & 0.04 & 0.92 & 0.92 \\
\bottomrule
\end{tabular}
\vspace{1em}

\textbf{(B) Spearman correlations with cumulative betamethasone dose}\\[0.3em]
\begin{tabular}{l rr r}
\toprule
Dimension & $\rho$ & $p$ & FDR $q$ \\
\midrule
FM\_Dim493 & +0.225 & 0.28 & 0.88 \\
FM\_Dim52 & +0.140 & 0.50 & 0.88 \\
FM\_Dim318 & -0.123 & 0.56 & 0.88 \\
FM\_Dim220 & +0.099 & 0.64 & 0.88 \\
FM\_Dim367 & +0.085 & 0.68 & 0.88 \\
FM\_Dim437 & -0.081 & 0.70 & 0.88 \\
FM\_Dim172 & +0.068 & 0.75 & 0.88 \\
FM\_Dim353 & +0.036 & 0.87 & 0.88 \\
FM\_Dim51 & -0.035 & 0.87 & 0.88 \\
FM\_Dim255 & +0.034 & 0.87 & 0.88 \\
FM\_Dim402 & -0.032 & 0.88 & 0.88 \\
\bottomrule
\end{tabular}
\vspace{1em}

\textbf{(C) Jonckheere--Terpstra trend tests}\\[0.3em]
\begin{tabular}{l rr l r}
\toprule
Dimension & $Z$ & $p$ & Trend & FDR $q$ \\
\midrule
FM\_Dim367 & -1.82 & 0.069 & decreasing & 0.58 \\
FM\_Dim255 & -1.46 & 0.14 & decreasing & 0.58 \\
FM\_Dim51 & -1.41 & 0.16 & decreasing & 0.58 \\
FM\_Dim318 & -0.86 & 0.39 & decreasing & 0.81 \\
FM\_Dim52 & 0.84 & 0.40 & increasing & 0.81 \\
FM\_Dim402 & -0.73 & 0.46 & decreasing & 0.81 \\
FM\_Dim493 & 0.54 & 0.59 & increasing & 0.81 \\
FM\_Dim353 & -0.45 & 0.65 & decreasing & 0.81 \\
FM\_Dim437 & -0.38 & 0.71 & decreasing & 0.81 \\
FM\_Dim220 & 0.34 & 0.74 & increasing & 0.81 \\
FM\_Dim172 & 0.13 & 0.90 & increasing & 0.90 \\
\bottomrule
\end{tabular}
}
\end{table}

\clearpage

\begin{table}[htbp]
\centering
\caption{Machine learning classification performance using complexity features. Stratified 5-fold group cross-validation. Chance level: 0.50 (binary), 0.33 (three-class).}
\label{tab:s5}
{\footnotesize
\textbf{(A) Binary classification (exposed vs.\ control)}\\[0.3em]
\begin{tabular}{ll cc c}
\toprule
Data & Classifier & Accuracy & AUC & F1 \\
\midrule
Unsmoothed & LogisticRegression & 0.413 +/- 0.087 & 0.430 +/- 0.116 & 0.437 +/- 0.089 \\
Unsmoothed & RandomForest & 0.438 +/- 0.051 & 0.444 +/- 0.080 & 0.484 +/- 0.062 \\
Unsmoothed & SVM & 0.410 +/- 0.025 & 0.512 +/- 0.149 & 0.456 +/- 0.054 \\
Unsmoothed & XGBoost & 0.447 +/- 0.072 & 0.430 +/- 0.083 & 0.478 +/- 0.064 \\
Smoothed & LogisticRegression & 0.437 +/- 0.074 & 0.426 +/- 0.113 & 0.467 +/- 0.097 \\
Smoothed & RandomForest & 0.449 +/- 0.048 & 0.422 +/- 0.043 & 0.497 +/- 0.059 \\
Smoothed & SVM & 0.374 +/- 0.037 & 0.548 +/- 0.135 & 0.425 +/- 0.066 \\
Smoothed & XGBoost & 0.451 +/- 0.024 & 0.438 +/- 0.039 & 0.484 +/- 0.020 \\
\bottomrule
\end{tabular}
\vspace{1em}

\textbf{(B) Three-class classification (low-dose, high-dose, control)}\\[0.3em]
\begin{tabular}{ll cc c}
\toprule
Data & Classifier & Accuracy & AUC & F1 (macro) \\
\midrule
Unsmoothed & LogisticRegression & 0.397 +/- 0.113 & 0.475 +/- 0.010 & 0.261 +/- 0.064 \\
Unsmoothed & RandomForest & 0.341 +/- 0.073 & 0.478 +/- 0.036 & 0.255 +/- 0.048 \\
Unsmoothed & SVM & 0.426 +/- 0.148 & 0.481 +/- 0.045 & 0.215 +/- 0.079 \\
Unsmoothed & XGBoost & 0.334 +/- 0.096 & 0.455 +/- 0.049 & 0.258 +/- 0.054 \\
Smoothed & LogisticRegression & 0.401 +/- 0.111 & 0.485 +/- 0.014 & 0.259 +/- 0.046 \\
Smoothed & RandomForest & 0.338 +/- 0.073 & 0.465 +/- 0.043 & 0.254 +/- 0.051 \\
Smoothed & SVM & 0.421 +/- 0.134 & 0.472 +/- 0.042 & 0.201 +/- 0.040 \\
Smoothed & XGBoost & 0.330 +/- 0.101 & 0.452 +/- 0.023 & 0.273 +/- 0.073 \\
\bottomrule
\end{tabular}
}
\end{table}

\begin{figure}[htbp]
\centering
\includegraphics[width=\textwidth]{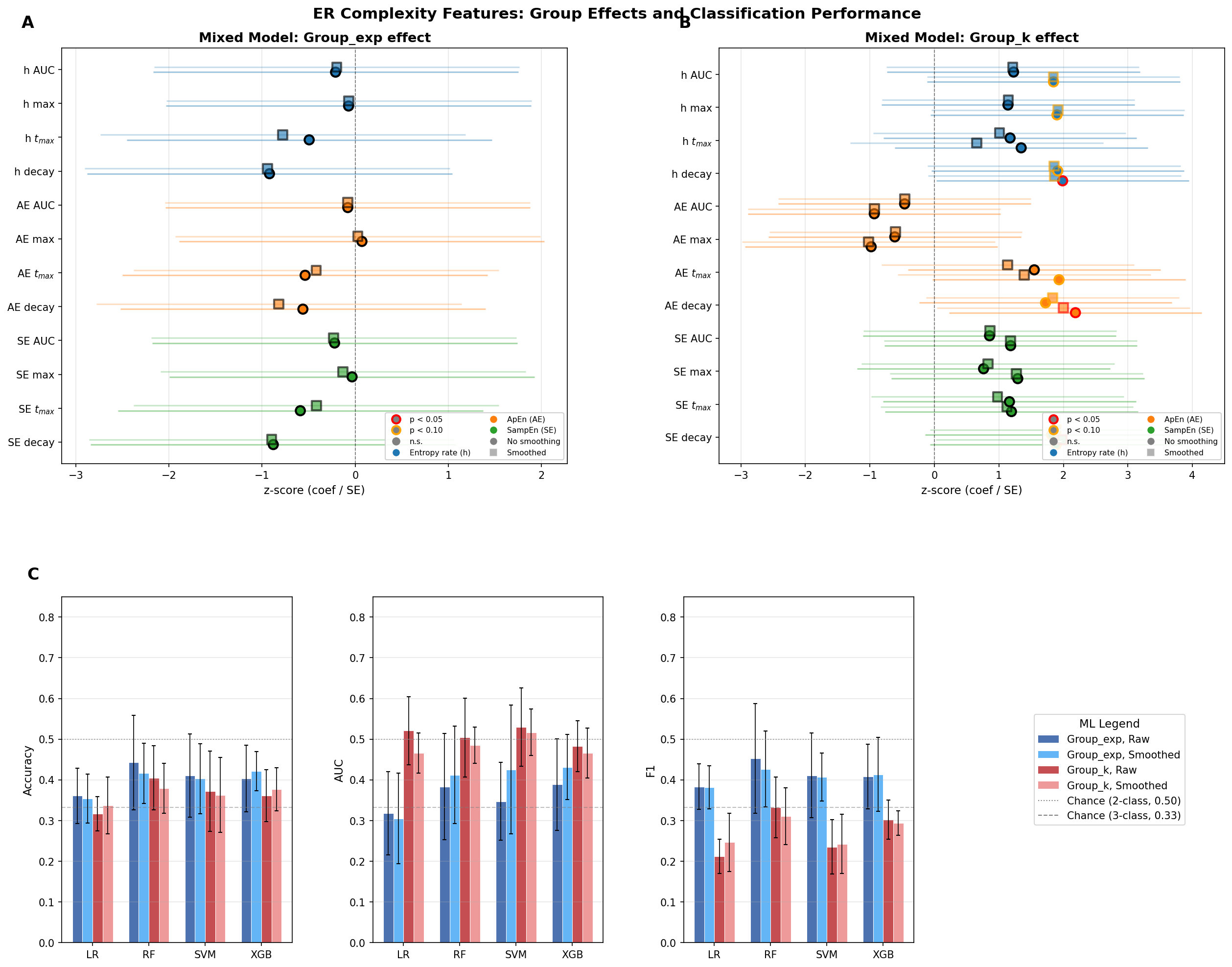}
\caption{Multi-panel summary of complexity analysis: forest plots of binary and dose-based $z$-scores, and ML classification performance.}
\label{fig:summary}
\end{figure}

\begin{figure}[htbp]
\centering
\includegraphics[width=\textwidth]{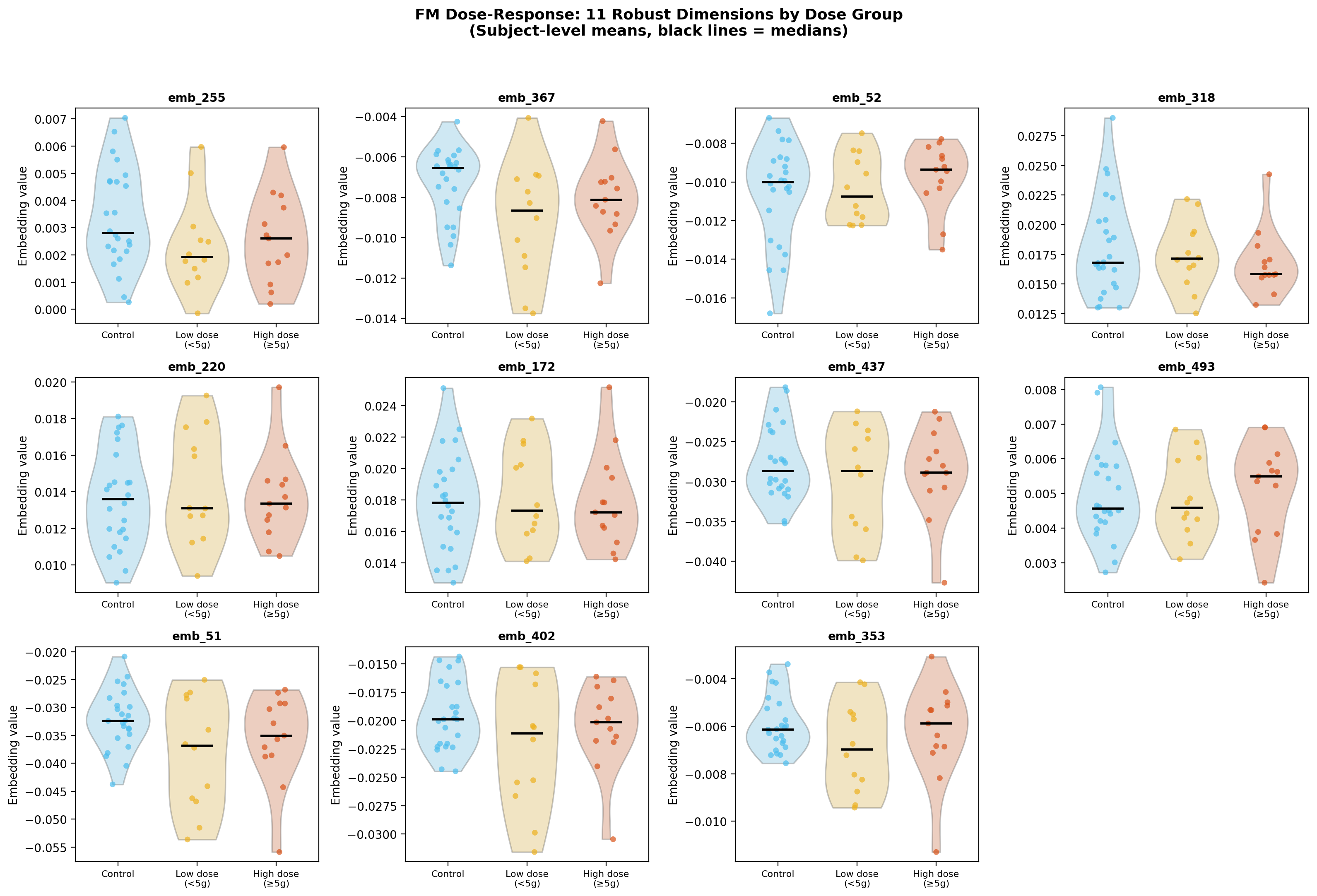}
\caption{Violin plots of all 11 FM dimensions by dose group (control, low-dose, high-dose), showing subject-level averages across segments.}
\label{fig:violin_all}
\end{figure}

\begin{figure}[htbp]
\centering
\includegraphics[width=\textwidth]{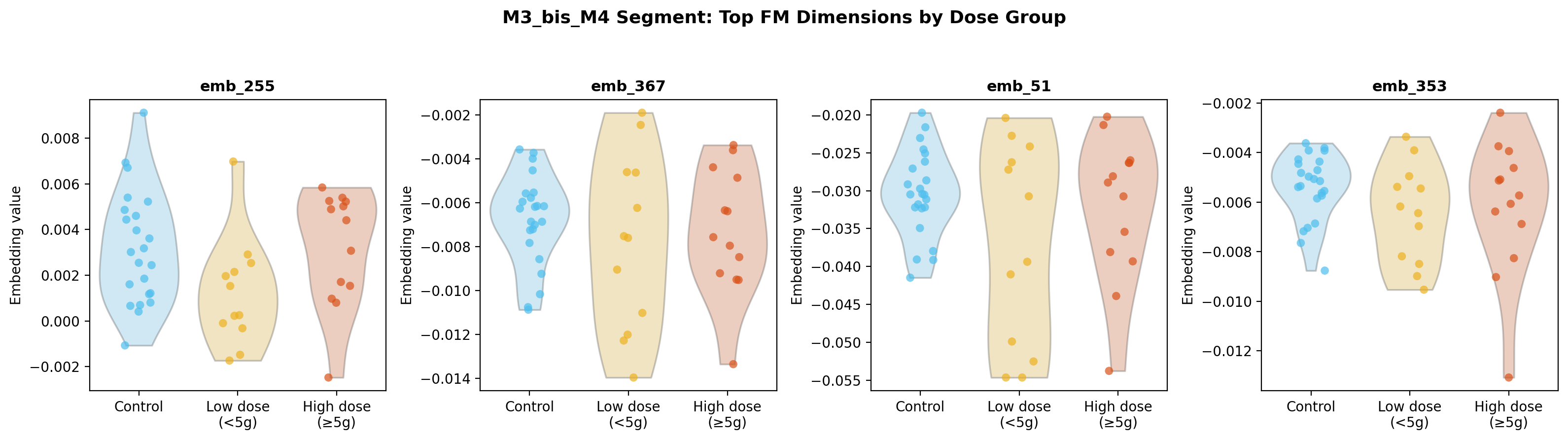}
\caption{Violin plots of top 4 FM dimensions in the mental arithmetic segment by dose group.}
\label{fig:violin_m3m4}
\end{figure}

\begin{figure}[htbp]
\centering
\includegraphics[width=\textwidth]{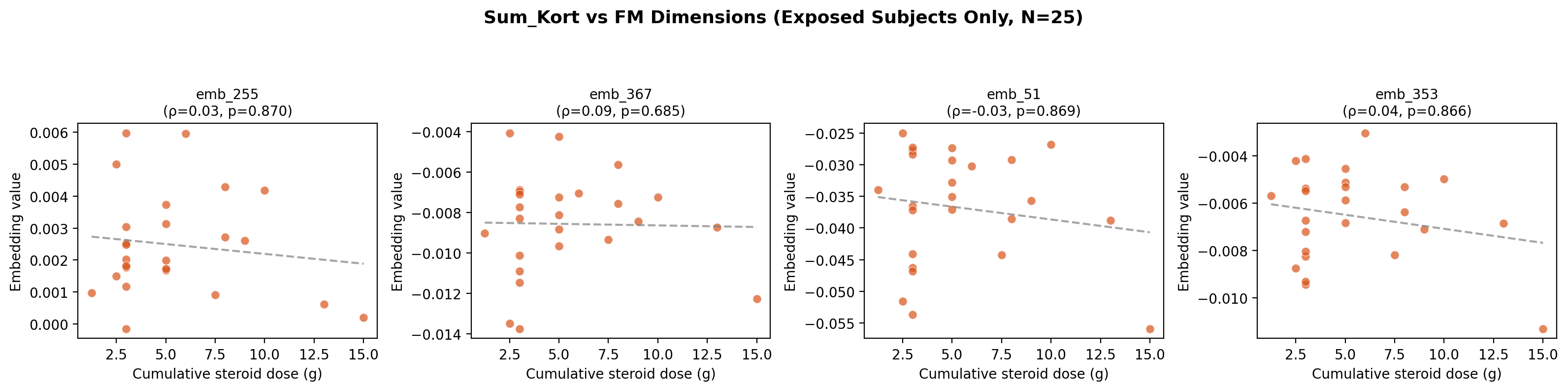}
\caption{Scatter plots of FM dimension values versus cumulative steroid dose for exposed subjects, with Spearman correlations.}
\label{fig:scatter}
\end{figure}

\end{document}